\documentclass[letter]{aa}
\usepackage{graphicx}
\usepackage[varg]{txfonts}
\usepackage{natbib}
\bibpunct{(}{)}{;}{a}{}{,} 
\makeatletter
\renewcommand*{\@fnsymbol}[1]{\ensuremath{\ifcase#1\or\star \or\dagger\fi}}
\makeatother
\begin{document}

\title{VLTI-GRAVITY measurements of cool evolved stars
\thanks{Based on observations made with the VLT Interferometer
at Paranal Observatory under programme IDs 60.A-9176 and 098.D-0647.}
}
\subtitle{I. Variable photosphere and extended atmosphere 
of the Mira star R Peg}
 \titlerunning{VLTI-GRAVITY measurements of R Peg}
\author{
M.~Wittkowski\inst{1}\and
G.~Rau\inst{2,3}\and
A.~Chiavassa\inst{4}\and
S.~H\"ofner\inst{5}\and
M.~Scholz\inst{6,7}\thanks{Deceased.}\and
P.~R.~Wood\inst{8}\and
W.~J.~de~Wit\inst{9}\and
F.~Eisenhauer\inst{10}\and
X.~Haubois\inst{9}\and
T.~Paumard\inst{11}
}
\institute{
European Southern Observatory, Karl-Schwarzschild-Str. 2,
85748 Garching bei M\"unchen, Germany,
\email{mwittkow@eso.org}
\and
NASA Goddard Space Flight Center, Code 667, Greenbelt, MD 20771, USA
\and
Department of Physics, The Catholic University of America,
Washington, DC 20064, USA
\and
Universit\'e C\^ote d’Azur, Observatoire de la C\^ote d’Azur, CNRS, 
Lagrange, CS 34229, 06304 Nice Cedex 4, France
\and
Division of Astronomy and Space Physics, Department of Physics and Astronomy, 
Uppsala University, Box 516, 75120 Uppsala, Sweden
\and
Zentrum f\"ur Astronomie der Universit\"at Heidelberg (ZAH),
Institut f\"ur Theoretische Astrophysik, Albert-Ueberle-Str. 2,
69120 Heidelberg, Germany
\and
Sydney Institute for Astronomy, School of Physics, University of Sydney,
Sydney NSW 2006, Australia
\and
Research School of Astronomy and Astrophysics, Australian National University, 
Canberra, ACT2611, Australia
\and
European Southern Observatory, Casilla 19001, Santiago 19, Chile
\and
Max Planck Institute for extraterrestrial Physics, Giessenbachstr., 
D-85748 Garching, Germany
\and
LESIA, Observatoire de Paris, Universit\'e PSL, CNRS, Sorbonne Universit\'e, 
Univ. Paris Diderot, Sorbonne Paris Cit\'e, 5 place Jules Janssen, 92195 Meudon, France
}
\date{Received \dots; accepted \dots}
\abstract{Dynamic model atmospheres of Mira stars predict
variabilities in the photospheric radius and in atmospheric molecular layers which are not yet strongly constrained by observations.}
{Here we  measure the variability
of the oxygen-rich Mira star R Peg
in near-continuum and molecular bands.
}
{We used near-infrared $K$-band spectro-interferometry with a 
spectral resolution of about 4000 obtained
at four epochs between post-maximum and minimum visual phases employing the 
newly available
GRAVITY beam combiner at the Very Large Telescope Interferometer (VLTI).}
{Our observations show a continuum radius that is anti-correlated with 
the visual lightcurve. Uniform disc (UD) angular diameters at a near-continuum
wavelength of 2.25\,$\mu$m are steadily increasing with values of 
8.7$\pm$0.1\,mas, 9.4$\pm$0.1\,mas, 
9.8$\pm$0.1\,mas, and 9.9$\pm$0.1\,mas at visual phases
of 0.15, 0.36, 0,45, 0.53, respectively. 
UD diameters at a bandpass around
2.05\,$\mu$m, dominated by water vapour,
follow the near-continuum
variability 
at larger UD diameters between 10.7\,mas and
11.7\,mas. UD diameters at the CO 2-0 bandhead, instead, are correlated with the
visual lightcurve and anti-correlated with the near-continuum UD diameters,
with values between 12.3\,mas and 11.7\,mas.}
{The observed anti-correlation between continuum radius and visual lightcurve
is consistent with an earlier  study of the oxygen-rich Mira S Lac, and
with recent 1D CODEX dynamic model atmosphere predictions. The amplitude
of the variation is comparable to the earlier observations of S Lac, and
smaller than predicted by CODEX models.
The wavelength-dependent
visibility variations at our epochs can be reproduced by a set of 
CODEX models at model phases between 0.3 and 0.6.
The anti-correlation
of water vapour and CO contributions at our epochs suggests
that these molecules undergo different processes
in the extended atmosphere along the stellar cycle.
The newly available GRAVITY instrument is suited to conducting longer  
time series observations, which are needed to 
provide strong constraints on the model-predicted 
intra- and inter-cycle variability.}
\maketitle
\section{Introduction}
\label{sec:intro}
Low- to intermediate-mass stars evolve to red giant and
asymptotic giant branch (AGB) stars. Mass loss increases
during the AGB evolution. 
The AGB mass loss is driven by an interplay between pulsations, which extend
the atmosphere, dust formation in the extended atmosphere, and radiation
pressure on the dust  \citep[e.g.][]{Wachter2002,Mattsson2010,Hoefner2018}. 
However, details of these interrelated processes are a matter of debate, 
in particular regarding the atmospheric levitation and wind acceleration 
for oxygen-rich stars \citep[e.g.][]{Bladh2015,Hoefner2016,Bladh2017}.
These processes are better understood for carbon-rich AGB stars 
\citep[e.g.][]{Nowotny2010,Nowotny2011,Rau2015,Rau2017}.

Dynamic 1D model atmospheres
of oxygen-rich Mira stars based on self-excited radial pulsation models 
(CODEX models) by \citet{Ireland2008,Ireland2011}
predicted a regular sinusoidal variation of the photospheric radius and
an irregular chaotic variability of the outer molecular layers. 
The latest 3D radiation hydrodynamic (RHD) simulations of AGB stars (CO5BOLD)
by \citet{Freytag2008} and \citet{Freytag2017} show non-radial structures
such as long-lasting giant convection cells and short-lived surface granules.
These dynamical phenomena trigger large-scale atmospheric shock waves that
expand roughly spherically  and are similar to those of 1D models, except
they do not cover the full surface at a given instance.

\citet{Thompson2002} conducted long-term narrow-band interferometric 
monitoring over 2--3 stellar cycles of the oxygen-rich Mira S Lac and the
carbon-rich Mira RZ Peg. Their data shows the expected sinusoidal variation
in the continuum angular radius and different phase lags in the continuum
and molecular bands. For the oxygen-rich Mira S Lac, the continuum minimum
size tracked the visual maximum brightness, i.e. the continuum size and
the visual lightcurve were anti-correlated. For the carbon-rich Mira RZ Peg, the
phase lag was 0.28.
A few other interferometric 
studies included fewer epochs and confirmed the presence of 
variability in the continuum size and in molecular layers
\citep[e.g.][]{Woodruff2009,Haubois2015,Wittkowski2016}.

Here we report on measurements of the variability in  the 
continuum radius and in  extended molecular layers using 
the example of the oxygen-rich Mira star R Peg obtained during science 
verification and early science operations of the newly available 
near-infrared $K$-band beam combiner GRAVITY \citep{Gravity2017} at 
the VLT Interferometer (VLTI).
\section{Observations and data reduction}
\label{sec:obs}
We used the GRAVITY instrument during science verification
and early science operation periods to perform high-precision multi-epoch
spectro-interferometry of the Mira variable R~Peg. 

R~Peg is an oxygen-rich Mira variable of spectral type M6--M9
with a mean period of 378\,d \citep{Samus2017}. 
We adopt the parallax to R~Peg from the Gaia DR 2 \citep{Gaia2016,Gaia2018}
of $\varpi$=2.8300$\pm$0.2544\,mas. The relative error of $f=0.09$
is quite small, so  we can simply estimate the distance by inverting 
the parallax
\citep{Luri2018}. This distance is $\rho$=353$^{+35}_{-29}$\,pc.
Figure~\ref{fig:lightcurve} shows the
recent visual lightcurve of R~Peg based on data obtained from the
AAVSO (American Association of Variable Star Observers) and
AFOEV (Association Francaise des Observateurs d'Etoiles Variables)
databases. 
Table~\ref{tab:obslog} provides the log of our observations.
We fit a sine curve to the ten most recent cycles of the visual
lightcurve relative to our observations (JDs 2453800--2457600) to estimate 
a current period of 377.1 $\pm$ 0.2\,d and to assign visual phases of
0.15, 0.36, 0.45, and 0.53 to our four epochs of observation in June, 
September, October, and November 2016.
R~Peg transitioned from post-maximum to minimum phases
during our four epochs. We used the compact baseline configuration for all
observations, where the 1.8\,m auxiliary telescopes (ATs) were located 
at stations A0, B2, C1, and D0, giving ground baseline lengths between 11.3\,m
and 33.9\,m, and projected baseline lengths between 9.8\,m and 32.0\,m.
We bracketed observations of R~Peg by observations 
of interferometric calibrators in sequences of calibrator-science-calibrator 
observations. Information on the calibrators and their adopted angular 
diameters is available in Tables \ref{tab:obslog} and ~\ref{tab:calibrators}.
The spectral range of our GRAVITY observations included
the full near-infrared $K$ band between 1.99\,$\mu$m and 2.45\,$\mu$m
at a spectral resolution of $R\sim4000$ (high spectral resolution mode). 
The GRAVITY instrument allows us to choose between splitting the light
into two polarization angles (split polarization mode), which increases 
the internal fringe contrast, or to use the combined light (combined 
polarization mode), which increases the sensitivity of the instrument. As our
target is bright and sensitivity was not an issue, we chose to use
the split polarization mode.
The detector integration time (DIT) on
the science spectrometer was 5\,s for the observations in June and
September, and 10\,s for those in October and November. On each night 
the same value was used for R~Peg and its calibrators.
The fringe tracker
was operated with a DIT of 0.0085 sec for all observations.

We reduced and calibrated the data with the latest release of the
GRAVITY pipeline (version 1.10.11)\footnote{Available at https://www.eso.org/sci/software/pipelines/gravity/gravity-pipe-recipes.html}  
and its Reflex workflows {\tt gravity-wk} and {\tt gravity-viscal}, 
respectively. Each observation was executed as a sequence
between object and sky positions (object-sky-object-sky), and we
computed averages of the object and sky files.
The pipeline gives results for each of the two polarization directions 
separately,
and we averaged the two results during post-processing. There were no
significant differences between them. Fringe tracker data, obtained
at a low spectral resolution of five spectral channels across the $K$-band, 
were processed in the same way.

As a reference, all calibrated visibility spectra are shown in 
Figs.~\ref{fig:visjun}--\ref{fig:visnov}, together with the model fits as
described below. The visibility 
spectra show the 
typical shapes of oxygen-rich Mira stars as observed previously with the 
VLTI-AMBER instrument \citep[e.g.][]{Wittkowski2011,Wittkowski2016}.
The visibility spectra show a maximum at the near-continuum bandpass
around 2.25\,$\mu$m, corresponding to a minimum angular diameter, 
where the bandpass is only slightly contaminated by molecular layers.
Toward shorter and longer wavelengths the visibility drops due to 
the presence of extended molecular layers; in the $K$ band most importantly it drops 
due to H$_2$O and CO.
The GRAVITY data show an increased precision with respect to the AMBER data; 
provide the data for six baselines in one snapshot, while the AMBER data provide three; and provide a spectral resolution of $\sim$4000 across the 
full $K$ band compared to a resolution of $\sim$1500 for AMBER across 
two separate halves of the $K$ band.
The closure phase spectra are not an essential part of the analysis
of this work which focuses on overall diameter variations
based on visibility data in the first lobe.
For reasons of completeness, we show the closure phase data in
Figs.~\ref{fig:cpjun}--\ref{fig:cpnov}.

The data obtained on the science spectrometer mostly agree 
well with those obtained on the fast fringe tracker, which 
confirms a high accuracy of the absolute visibility calibration. In a few
visibility spectra from September and October, the fringe tracker results
deviate from the science spectrometer results for some of the baselines, 
especially toward the red part of the spectrum for some baselines. 
Based on the shape of the visibility spectra and the consistency of the 
science camera results across different baselines, we attribute these
differences to a problem with the {absolute} calibration of the fringe 
tracker data at these epochs, which does not affect the quality of the 
fringe tracking itself and whose origin is not yet known.
Some visibility spectra show a dip centred
at 2.00\,$\mu$m, close to the
blue edge of the wavelength range, which is an artefact and most likely caused 
by low flux at these wavelengths owing to absorption by water, or due to
a contamination from the metrology laser operating at 1.91\,$\mu$m.
\section{Data analysis}
\label{sec:analysis}
\begin{table}
\caption{\label{tab:fitud} Uniform disc fit results }
\centering
\begin{tabular}{lrrrr}
\hline\hline
Band          & \multicolumn{4}{c}{UD diameter} \\
($\mu$m)      & \multicolumn{4}{c}{(mas)}       \\
              & Jun/   &  Sep/  &   Oct/ &   Nov/  \\\
       & $\phi$=0.15  & $\phi$=0.36  &  $\phi$=0.45 &  $\phi$=0.53  \\\hline
2.25 (cont.)  &  8.67 &  9.42 &  9.76 &  9.89  \\
2.05 (water)  & 10.73 & 11.27 &  11.60&  11.73 \\
2.29 (CO 2-0) & 12.27 & 11.70 & 11.64 & 11.67  \\
2.32 (CO 3-1) & 12.77 & 11.76 & 10.98 & 12.18  \\
\hline
\end{tabular}
\tablefoot{We adopt errors of the UD diameters of 1\%.}
\end{table}
\begin{figure}
\resizebox{\hsize}{!}{\includegraphics{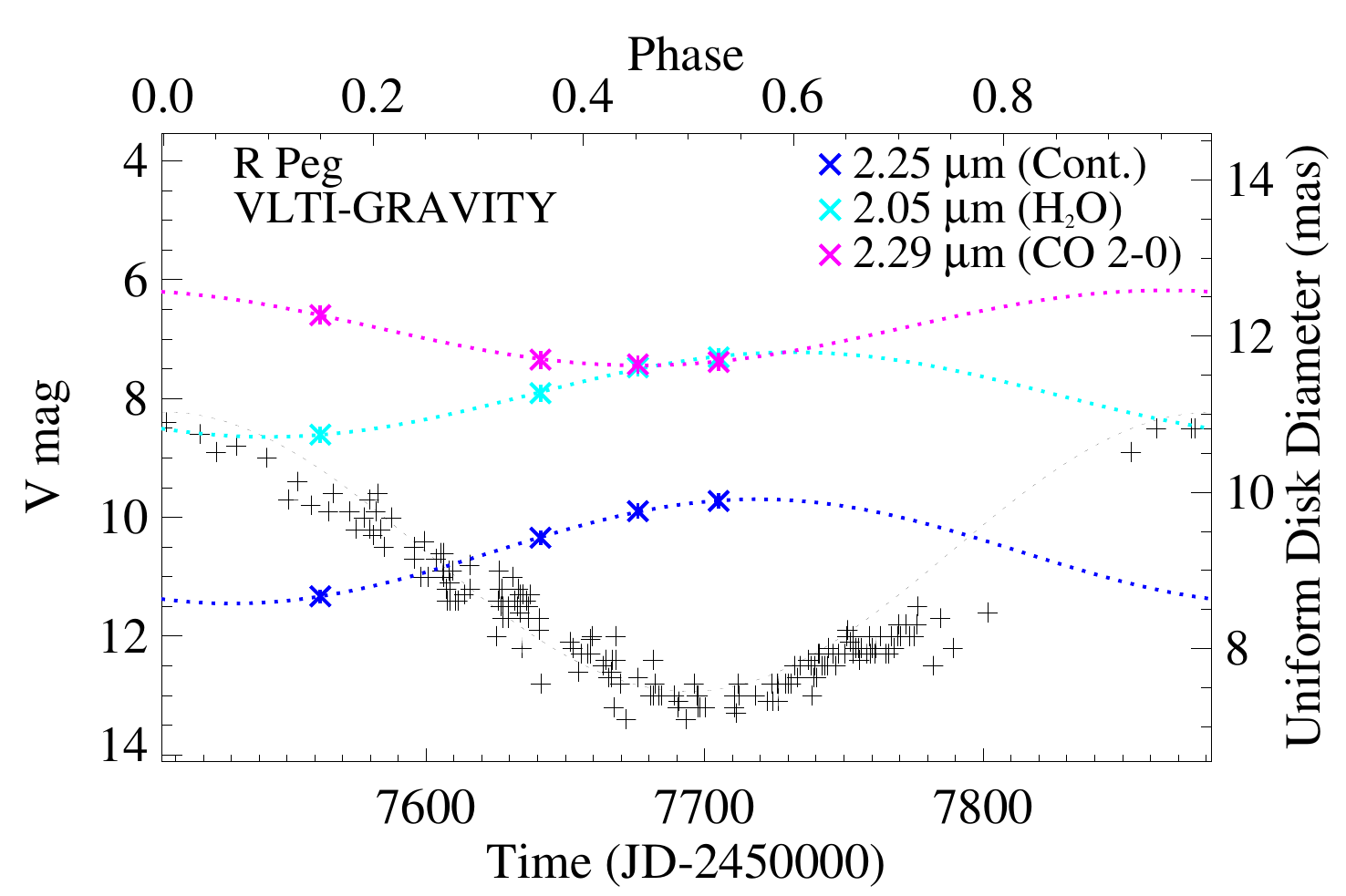}}
\caption{Variability in the $V$ magnitude and in the UD angular diameters
in the near-continuum band, and in bands dominated by H$_2$O (2.05\,$\mu$m)
and CO 2-0 (2.29\,$\mu$m). Also shown are sinusoidal fit results. The
minimum continuum size tracks the maximum light, which can be understood by 
the increasing effective temperature when the star gets smaller. The minimum
contribution of H$_2$O tracks the maximum light as well, which can be
understood as the destruction of water vapour at maximum light and formation
at minimum light. The contribution from CO instead is greatest at maximum
light, indicating a different and more stable  behaviour of CO compared to 
water vapour.}
\label{fig:diameters}
\end{figure}
\begin{table}
\caption{\label{tab:sinusoidal} Sinusoidal fit results}
\centering
\begin{tabular}{lrrrr}
\hline\hline
Value    & Last max. & Phase & Mean  & Full \\
      &  (d)\tablefootmark{a}   & lag      & Value            & Ampl.          \\\hline
$V$ mag        &  7505       & --         &  10.58 & 4.71      \\
$\Theta_\mathrm{Cont.}$ (mas) &  7719       & 0.57      &  9.24 & 1.33      \\
$\Theta_\mathrm{H_2O}$ (mas) &  7732       & 0.60      &  11.25& 1.08      \\
$\Theta_\mathrm{CO (2-0)}$ (mas) &  7488       & -0.05     &  12.11& 0.96      \\
$\Theta_\mathrm{CO (3-1)}$ (mas) &  7468       & -0.10     &  12.78& 2.09      \\
\hline
\end{tabular}
\tablefoot{
\tablefootmark{a}{JD-2 450 000.}
We adopt a period of 377 days for all fits, based on the fit to the last 
ten periods of the
$V$ magnitude, see Fig.~\protect\ref{fig:lightcurve}.}
\end{table}
We are interested in the changes of the photospheric radius 
and that of extended molecular layers across our four epochs transitioning
from post-maximum to minimum visual phases.
As a model-independent analysis, we first provide fits of a uniform disc (UD)
model at a few selected bandpasses before  discussing
a comparison 
of our data with dynamic model atmosphere predictions.
The intensity profile at the photospheric layer in a near-continuum
bandpass is expected to be well described by a 
UD \citep[e.g.][Fig. 3b]{Wittkowski2016}, so that we
chose this simple geometrical model. The intensity profile in bandpasses
corresponding to molecular layers may be more complex, and may typically
resemble geometrical profiles with two  or more components. With our strategy
of snapshot observations at different epochs, we are not able to constrain
the true shape of the intensity profile at these wavelengths. This would 
require time series of image reconstructions, which is much more expensive 
in terms of observing time. Nevertheless, the deviation of a uniform disc
diameter in a molecular band compared to that in a near-continuum band
can be used as a simple measure of the overall contribution of the molecular
layer to the intensity profile, which includes effects of both its radial 
extension and its relative intensity level. Followed over different epochs, 
the UD analysis gives information about the time variability of the 
molecular contribution.
We chose observations in the first lobe of the
visibility function which are sensitive to first-order structure, i.e.
the overall radial extent, and much less sensitive to higher order structures such as the 
the limb-darkening effect or even surface inhomogeneities caused by 
convection cells.
\subsection{Uniform disc angular diameters}
We obtained at each epoch UD diameters at the near-continuum bandpass
at 2.25\,$\mu$m to characterize the photospheric continuum radius. 
The bandpass at 2.25\,$\mu$m is expected to be almost free of 
molecular contamination for oxygen-rich Mira stars
\citep[][Fig. 4]{Wittkowski2008}. 
We chose a bandpass at 2.05\,$\mu$m to monitor the strength of water 
vapour layers, excluding the instrumental artificial dip in some 
visibility functions at 2.00\,$\mu$m mentioned above.
Finally, we chose narrow bandpasses  at the lowest points of the visibility 
drop in the CO (2-0) bandhead at 2.29\,$\mu$m and in the CO (3-1) bandhead
at 2.32\,$\mu$m.

Table~\ref{tab:fitud} shows the UD fit results at these bandpasses for
each of our epochs. We adopt errors of $\sim$1\% including
calibration uncertainties, while the formal errors are much lower.
The synthetic visibility values of the UD fits are represented by
blue dots in Figs.~\ref{fig:visjun}--\ref{fig:cpnov}. They show 
a good consistency between the visibility spectra corresponding to the
different baselines, which again demonstrates the high accuracy of the
absolute visibility calibration. Figure~\ref{fig:diameters} plots the
resulting UD diameters compared to the contemporaneous visual lightcurve,
together with sinusoidal fits, as described below.

Uniform disc  angular diameters at the near-continuum bandpass
 steadily increase with values of 8.7$\pm$0.1\,mas, 9.4$\pm$0.1\,mas,
9.8$\pm$0.1\,mas, and 9.9$\pm$0.1\,mas from post-maximum to minimum 
visual phases of 0.15, 0.36, 0,45, and 0.53, respectively. This means that
the continuum radius is anti-correlated to the visual lightcurve across our 
epochs. This
behaviour is consistent with the multi-epoch narrow-band interferometric 
study of the oxygen-rich Mira S Lac by \citet{Thompson2002}, 
which -- to our knowledge -- represents the only other such study 
in the literature. 

Uniform disc diameters at the water vapour bandpass around
2.05\,$\mu$m follow the variability of the near-continuum radii at larger 
UD diameters between 10.7\,mas and 11.7\,mas. 
UD diameters at the CO 2-0 bandhead, instead, are correlated with the
visual lightcurve and anti-correlated with the near-continuum UD diameters
with values between 12.3\,mas and 11.7\,mas. 

We computed best-fit sinusoidal functions to the UD diameters in the 
different bandpasses.
The results are listed
in Table~\ref{tab:sinusoidal}  and plotted in Fig.~\ref{fig:diameters}. 
Although the measured points lie well on the sinusoidal curves, the fit
is not well constrained due to the low number of epochs. More epochs are
needed to study the intra- and inter-cycle variability in more detail.
Nevertheless, this fit provides us with a characterization of the phase lags 
with respect to the visual lightcurves, and with estimates
of the mean values and amplitudes of the diameter and lightcurve variation.
The continuum diameter lags behind the visual lightcurve by 0.57. The water vapour
diameter similarly lags behind the visual lightcurve  by 0.60. The CO (2-0) and 
CO (2-0) diameters have different phase lags of -0.05 and 0.10. 
\citet{Thompson2002} derived
similar phase lags for S Lac of 0.48 in the continuum and 0.53 in a 
water vapour bandpass at 2.0\,$\mu$m. Their bandpass at 2.4\,$\mu$m 
had a phase lag of 0.34, but is  mostly dominated by water vapour
as well, and may not track the CO bandheads.

Phenomenologically, our observed phase lags of sizes or contributions
in continuum, water vapour, and CO bands with respect to the visual 
lightcurve can be interpreted as follows. When the continuum radius
decreases, the effective temperature rises, and this increase in
effective temperature dominates the lightcurve ($L\propto R T_\mathrm{eff}^4$).
At the same time, water vapour is an unstable molecule that can get
destroyed at maximum effective temperature and luminosity, and can be formed 
at minimum effective temperature and luminosity 
\citep[e.g. Fig. 9 in][]{Bladh2013};  its contribution
is thus anti-correlated with the lightcurve and correlated with the continuum 
size. CO instead is a more stable 
molecule whose abundance may be more stable across the stellar cycle,
and which may become more strongly excited
at maximum light than at minimum light. Time series of images in water and CO 
bands may help to further investigate the different behaviour of these 
molecules as a function of visual phase, and to disentangle variations
in geometrical size and in flux contribution.
\subsection{Fundamental stellar parameters}
\label{sec:fundamental}
\begin{table}
\centering
\caption{Mean fundamental parameters of R~Peg. }
\label{tab:fundpar}
\begin{tabular}{llr}
\hline\hline
Parameter                               & Value            & Ref. \\\hline
Period (d)                               & 377              & 1 \\
Distance (pc)                           & 353$^{+35}_{-29}$& 2 \\
Interstellar extinction $(A_V$)           & 0.20             & 3 \\
Bolometric flux (10$^{-9}$ W/m$^2$)     & 1.08 $\pm$ 0.05  & 4 \\
Rosseland angular diameter (mas)        & 9.24 $\pm$ 0.09  & 5 \\
Rosseland radius ($R_\sun$)             & 351$^{+38}_{-31}$& 6 \\
Effective temperature (K)               & 2480 $\pm$ 40    & 7 \\
Luminosity ($\log L/L_\sun$)           & 3.62 $\pm$ 0.12  & 8 \\
Mass ($M_\sun$)                         & 1.0  $\pm$ 0.3   & 9\\
Surface gravity ($\log g$)              & -0.65 $\pm$ 0.24 & 10 \\\hline
\end{tabular}
\tablefoot{
1: Analysis of the AAVSO/AFOEV lightcurve, see Sect.~\ref{sec:obs} 
and Fig.~\ref{fig:lightcurve};
2: Based on the parallax from Gaia DR 2, see Sect.~\ref{sec:obs}; 
3: \citet{Whitelock2008};
4: integrated photometry, see Sect.~\ref{sec:fundamental};
5: this work, mean continuum angular radius from Table~\ref{tab:sinusoidal};
6: calculated from 2 and 5;
7: calculated from 4 and 5;
8: calculated from 2 and 4;
9: assumed mass for an oxygen-rich Mira;
10: calculated from 6 and 9.
}
\end{table}

We used the mean continuum angular diameter of 9.24$\pm$0.09\,mas together
with our adopted distance and the bolometric flux to derive the mean radius,
effective temperature, and luminosity of R~Peg. We estimated a bolometric
flux of 1.08$\pm$0.05\,10$^{-9}$\,W/m$^2$ using broad-band 
$UBVRIJHK$ \citep{Ducati2002} and IRAS 
\citep{iras} photometry, and using an $A_V$ value of 
0.20\,mag from \citet{Whitelock2008}. 
We obtained a mean continuum radius of
353$^{+35}_{-29}$\,R$_\odot$, a mean effective temperature of 2480$\pm$40\,K,
and a mean luminosity of 4200$^{+1320}_{-960}$\,L$_\odot$. Assuming a mass of 
1$\pm$0.3\,M$_\odot$, this corresponds to a surface gravity of -0.65$\pm$0.25.
Table~\ref{tab:fundpar} provides an overview of these stellar parameters.
\subsection{Comparison to one-dimensional CODEX dynamic model atmospheres}
\begin{table}
\caption{\label{tab:fitcodex} CODEX fit results }
\centering
\begin{tabular}{lrrrr}
\hline\hline
                             & Jun/         &  Sep/        &   Oct/       &   Nov/        \\
                             & $\phi$=0.15  & $\phi$=0.36  &  $\phi$=0.45 &  $\phi$=0.53  \\\hline
Series                       & o54          & o54          &  o54         &  o54          \\
Model                        & 261460       & 261740       &  261940      &  261940       \\
$\phi_\mathrm{Model}$        & 0.31         & 0.40         &  0.60        &  0.60         \\
$\Theta_\mathrm{Cont}$ (mas) & 8.82         & 8.95         &  9.15        &  9.30         \\
\hline
\end{tabular}
\tablefoot{We adopt errors of the diameters of 1\%.}
\end{table}
\citet{Ireland2008,Ireland2011} presented dynamic model
atmospheres for Mira stars that are based on self-excited pulsation 
models, and use the opacity sampling method to represent gas opacities. 
CODEX models are available in four different series
that correspond to different stellar parameters of the underlying
hypothetical non-pulsating parent star, covering radii between
209\,R$_\odot$ and 201\,R$_\odot$, luminosities between 5050\,L$_\odot$
and 8160\,L$_\odot$, effective temperatures between 2860\,K and 3400\,K,
and periods between 307\,d and 427\,d, meant to represent the Miras o~Cet,
R Leo, and R Cas. R~Peg has a larger radius and lower effective temperature
 compared to the parameters of the CODEX model series. 
However, general properties of the variability of the CODEX models may 
be compared to our observations.

The CODEX models predicted in general a regular sinusoidal variation in the 
photospheric radius and a more irregular variability in the outer molecular 
layers. We inspected the variabilities of the luminosity and the continuum radii 
as a function of model phase based on the tables provided 
by \citet{Ireland2011}. The CODEX model phases match the visual lightcurves
within 0.1. The CODEX model series show the highest luminosities
at model phases between -0.1 and 0.2, and the lowest luminosities at model
phases between 0.54 and 0.70, depending on the model series.
They show the smallest continuum radii at model phases -0.1 and the largest 
continuum radii at model phases 0.3 to 0.6. 

As a result, it is a typical characteristic of the CODEX models that the 
continuum radius increases between maximum and minimum light, when the 
luminosity decreases. This is consistent with our observations of R~Peg
and with the observations of S~Lac by \citet{Thompson2002}, which both
showed an anti-correlation between the visual lightcurve and the continuum
radius. A closer inspection showed that the amplitudes of the radius 
variations of the CODEX models lie between 45\% and 67\%, while we estimated 
a clearly lower observed amplitude of 14\% (Tab.~\ref{tab:sinusoidal}). 
For comparison, the S Lac
observations by \citet{Thompson2002} showed an amplitude of 19\%.
Moreover, the shape of the radius variation in the CODEX models is steeper
than the observations. The model amplitude varies from pulsation cycle 
to pulsation cycle;   most of this variation occurs near minimum light 
when the models are most extended and coolest, and hence most uncertain 
\citep[e.g. Figs. 11--13 of][]{Ireland2011}.
However, even in the more regular 
part of the radius variation (phases about 0--0.25), the model amplitude 
is still larger than is the observed variation in R Peg. 
It is not clear whether these differences
between models and observations are caused by the different stellar 
parameters, or whether they indicate intrinsic differences in the models
compared to observations.

As a result of these differences between observations and models, 
we cannot find a set of CODEX models that reproduces well the details of the 
observed variability of the continuum radius. We made an attempt to find a
set of CODEX models that fits our observations  by treating the model
continuum diameter as a free parameter for each observation individually, 
effectively scaling the amplitude of the model variations to the observed
ones. We selected a set of models that (1) matches as closely as possible
the post-maximum to minimum visual phases of the observed epochs, and (2)
reproduces the observed wavelength dependence of the visibility spectra
and its variability.  Table~\ref{tab:fitcodex} lists the parameters of the 
identified CODEX model set that gives the best compromise between these
criteria. Figures~\ref{fig:visjun}--\ref{fig:cpnov} show
the synthetic visibility spectra compared to the observed spectra. 
This set of models consists of adjacent post-maximum to minimum
visual model phases as observed, reproduces the increasing continuum diameter
at these phases, and provides a satisfactory fit to the 
the wavelength dependent shape of the visibility spectra at the different
epochs.
The largest discrepancy between observations and model
is seen for epoch 1, where the model phase (0.3) differs most with respect
to the observed phase (0.15), and where the model still predicts 
deeper CO bandheads than observed.
\subsection{Comparison to three-dimensional CO5BOLD model atmospheres}
Recently, \citet{Freytag2017} presented 3D radiation-hydrodynamics models
of AGB stars. They used different model parameters covering radii between
294\,R$_\odot$ and 531\,R$_\odot$, luminosities between 4978\,L$_\odot$
and 10028\,L$_\odot$, effective temperature between 2506\,K and 2827\,K,
and periods between 338\,d and 820\,d. 
Models {\tt st28gm05n001} or {\tt st29gm06n001} from Table~1 of \citet{Freytag2017} 
are closest to 
the estimated stellar parameters of R Peg. 
We inspected variations in luminosity and radius of these 3D models, and 
found that the variability of both quantities is significantly more 
irregular than for the 1D CODEX models. This behaviour 
is expected because of the presence of additional non-radial structures 
such as long-lasting giant convection cells and short-lived surface granules. 
The model includes episodes where luminosity and radius are correlated,
and  episodes where these quantities are anti-correlated.
The maximum full amplitude of the radius variations in these models
is 14\% and 8\% for models {\tt st28gm05n001} or {\tt st29gm06n001}, respectively, 
while it is lower during individual cycles. These numbers are comparable to 
our observed radius variation of R~Peg of 14\%.
We chose not to attempt a direct comparison between our observations and 
predictions by 3D models  because our present observations cover only 
four epochs and the 3D models show this strong irregularity. Observations 
at more phases are needed to attempt a meaningful comparison of observations 
with 3D model predictions of the stellar variability.
\section{Summary and conclusions}
We have reported on measurements of the variability in the
continuum radius and in extended molecular layers at four epochs
between post-maximum and minimum visual phases for the example of the 
oxygen-rich Mira star R Peg.
To the best of our knowledge, our study represents the first characterization
of the phase lags between Mira angular sizes in continuum and molecular
bands and the lightcurve since the pioneering observations in this field by 
\citet{Thompson2002}. We show that the continuum size and the size in a 
bandpass that is dominated by water vapour are anti-correlated with the 
visual lightcurve. This behaviour is consistent with the result by
\citet{Thompson2002} for S Lac and with predictions by CODEX dynamic
model atmospheres. The size in the CO (2-0) instead follows 
the visual lightcurve more closely, indicating a different and more stable - behaviour
of CO compared to that of  water vapour.
The full amplitude of the near-continuum diameter variation of R Peg
of 14\% is comparable to the 19\% for  S Lac    measured 
by \citet{Thompson2002}. This amplitude is smaller than predicted
by CODEX model atmospheres (45\%--67\%), and closer to those
predicted by  3D RHD simulations by \citet{Freytag2017} of up to 14\%.
The wavelength-dependent visibility variations at our epochs can be 
reproduced by a set of
CODEX models at post-maximum to minimum visual model phases between 
0.3 and 0.6.
We show that the newly available GRAVITY instrument is 
well suited to conducting such high-precision time series of stellar sizes.
Longer time series of these observations, and eventually of interferometric
imaging, are needed to provide strong constraints on the 
model-predicted intra- and inter-cycle variability, and the interplay between
pulsation, convection, and mass loss.
\bibliographystyle{aa}
\bibliography{AGB-GRAVITY}

\begin{thebibliography}{30}
\expandafter\ifx\csname natexlab\endcsname\relax\def\natexlab#1{#1}\fi

\bibitem[{{Bladh} {et~al.}(2015){Bladh}, {H{\"o}fner}, {Aringer}, \&
  {Eriksson}}]{Bladh2015}
{Bladh}, S., {H{\"o}fner}, S., {Aringer}, B., \& {Eriksson}, K. 2015, \aap,
  575, A105

\bibitem[{{Bladh} {et~al.}(2013){Bladh}, {H{\"o}fner}, {Nowotny}, {Aringer}, \&
  {Eriksson}}]{Bladh2013}
{Bladh}, S., {H{\"o}fner}, S., {Nowotny}, W., {Aringer}, B., \& {Eriksson}, K.
  2013, \aap, 553, A20

\bibitem[{{Bladh} {et~al.}(2017){Bladh}, {Paladini}, {H{\"o}fner}, \&
  {Aringer}}]{Bladh2017}
{Bladh}, S., {Paladini}, C., {H{\"o}fner}, S., \& {Aringer}, B. 2017, \aap,
  607, A27

\bibitem[{{Chelli} {et~al.}(2016){Chelli}, {Duvert}, {Bourg{\`e}s}, {Mella},
  {Lafrasse}, {Bonneau}, \& {Chesneau}}]{Chelli2016}
{Chelli}, A., {Duvert}, G., {Bourg{\`e}s}, L., {et~al.} 2016, \aap, 589, A112

\bibitem[{{Ducati}(2002)}]{Ducati2002}
{Ducati}, J.~R. 2002, VizieR Online Data Catalog, 2237

\bibitem[{{Freytag} \& {H{\"o}fner}(2008)}]{Freytag2008}
{Freytag}, B. \& {H{\"o}fner}, S. 2008, \aap, 483, 571

\bibitem[{{Freytag} {et~al.}(2017){Freytag}, {Liljegren}, \&
  {H{\"o}fner}}]{Freytag2017}
{Freytag}, B., {Liljegren}, S., \& {H{\"o}fner}, S. 2017, \aap, 600, A137

\bibitem[{{Gaia Collaboration} {et~al.}(2018){Gaia Collaboration}, {Brown},
  {Vallenari}, {Prusti}, {de Bruijne}, {Babusiaux}, \&
  {Bailer-Jones}}]{Gaia2018}
{Gaia Collaboration}, {Brown}, A.~G.~A., {Vallenari}, A., {et~al.} 2018, ArXiv
  e-prints

\bibitem[{{Gaia Collaboration} {et~al.}(2016){Gaia Collaboration}, {Prusti},
  {de Bruijne}, {Brown}, {Vallenari}, {Babusiaux}, {Bailer-Jones}, {Bastian},
  {Biermann}, {Evans}, \& et~al.}]{Gaia2016}
{Gaia Collaboration}, {Prusti}, T., {de Bruijne}, J.~H.~J., {et~al.} 2016,
  \aap, 595, A1

\bibitem[{{Gravity Collaboration} {et~al.}(2017){Gravity Collaboration},
  {Abuter}, {Accardo}, {Amorim}, {Anugu}, {{\'A}vila}, {Azouaoui}, {Benisty},
  {Berger}, {Blind}, {Bonnet}, {Bourget}, {Brandner}, {Brast}, {Buron},
  {Burtscher}, {Cassaing}, {Chapron}, {Choquet}, {Cl{\'e}net}, {Collin},
  {Coud{\'e} Du Foresto}, {de Wit}, {de Zeeuw}, {Deen},
  {Delplancke-Str{\"o}bele}, {Dembet}, {Derie}, {Dexter}, {Duvert}, {Ebert},
  {Eckart}, {Eisenhauer}, {Esselborn}, {F{\'e}dou}, {Finger}, {Garcia}, {Garcia
  Dabo}, {Garcia Lopez}, {Gendron}, {Genzel}, {Gillessen}, {Gonte}, {Gordo},
  {Grould}, {Gr{\"o}zinger}, {Guieu}, {Haguenauer}, {Hans}, {Haubois}, {Haug},
  {Haussmann}, {Henning}, {Hippler}, {Horrobin}, {Huber}, {Hubert}, {Hubin},
  {Hummel}, {Jakob}, {Janssen}, {Jochum}, {Jocou}, {Kaufer}, {Kellner},
  {Kendrew}, {Kern}, {Kervella}, {Kiekebusch}, {Klein}, {Kok}, {Kolb}, {Kulas},
  {Lacour}, {Lapeyr{\`e}re}, {Lazareff}, {Le Bouquin}, {L{\`e}na}, {Lenzen},
  {L{\'e}v{\^e}que}, {Lippa}, {Magnard}, {Mehrgan}, {Mellein}, {M{\'e}rand},
  {Moreno-Ventas}, {Moulin}, {M{\"u}ller}, {M{\"u}ller}, {Neumann}, {Oberti},
  {Ott}, {Pallanca}, {Panduro}, {Pasquini}, {Paumard}, {Percheron}, {Perraut},
  {Perrin}, {Pfl{\"u}ger}, {Pfuhl}, {Phan Duc}, {Plewa}, {Popovic}, {Rabien},
  {Ram{\'{\i}}rez}, {Ramos}, {Rau}, {Riquelme}, {Rohloff}, {Rousset},
  {Sanchez-Bermudez}, {Scheithauer}, {Sch{\"o}ller}, {Schuhler}, {Spyromilio},
  {Straubmeier}, {Sturm}, {Suarez}, {Tristram}, {Ventura}, {Vincent},
  {Waisberg}, {Wank}, {Weber}, {Wieprecht}, {Wiest}, {Wiezorrek}, {Wittkowski},
  {Woillez}, {Wolff}, {Yazici}, {Ziegler}, \& {Zins}}]{Gravity2017}
{Gravity Collaboration}, {Abuter}, R., {Accardo}, M., {et~al.} 2017, \aap, 602,
  A94

\bibitem[{{Haubois} {et~al.}(2015){Haubois}, {Wittkowski}, {Perrin},
  {Kervella}, {M{\'e}rand}, {Thi{\'e}baut}, {Ridgway}, {Ireland}, \&
  {Scholz}}]{Haubois2015}
{Haubois}, X., {Wittkowski}, M., {Perrin}, G., {et~al.} 2015, \aap, 582, A71

\bibitem[{{Helou} \& {Walker}(1988)}]{iras}
{Helou}, G. \& {Walker}, D.~W., eds. 1988, {Infrared astronomical satellite
  (IRAS) catalogs and atlases. Volume 7: The small scale structure catalog},
  Vol.~7, 1--265

\bibitem[{{H{\"o}fner} {et~al.}(2016){H{\"o}fner}, {Bladh}, {Aringer}, \&
  {Ahuja}}]{Hoefner2016}
{H{\"o}fner}, S., {Bladh}, S., {Aringer}, B., \& {Ahuja}, R. 2016, \aap, 594,
  A108

\bibitem[{{H{\"o}fner} \& {Olofsson}(2018)}]{Hoefner2018}
{H{\"o}fner}, S. \& {Olofsson}, H. 2018, \aapr, 26, 1

\bibitem[{{Ireland} {et~al.}(2008){Ireland}, {Scholz}, \& {Wood}}]{Ireland2008}
{Ireland}, M.~J., {Scholz}, M., \& {Wood}, P.~R. 2008, \mnras, 391, 1994

\bibitem[{{Ireland} {et~al.}(2011){Ireland}, {Scholz}, \& {Wood}}]{Ireland2011}
{Ireland}, M.~J., {Scholz}, M., \& {Wood}, P.~R. 2011, \mnras, 418, 114

\bibitem[{{Luri} {et~al.}(2018){Luri}, {Brown}, {Sarro}, {Arenou},
  {Bailer-Jones}, {Castro-Ginard}, {de Bruijne}, {Prusti}, {Babusiaux}, \&
  {Delgado}}]{Luri2018}
{Luri}, X., {Brown}, A.~G.~A., {Sarro}, L.~M., {et~al.} 2018, ArXiv e-prints

\bibitem[{{Mattsson} {et~al.}(2010){Mattsson}, {Wahlin}, \&
  {H{\"o}fner}}]{Mattsson2010}
{Mattsson}, L., {Wahlin}, R., \& {H{\"o}fner}, S. 2010, \aap, 509, A14

\bibitem[{{Nowotny} {et~al.}(2011){Nowotny}, {Aringer}, {H{\"o}fner}, \&
  {Lederer}}]{Nowotny2011}
{Nowotny}, W., {Aringer}, B., {H{\"o}fner}, S., \& {Lederer}, M.~T. 2011, \aap,
  529, A129

\bibitem[{{Nowotny} {et~al.}(2010){Nowotny}, {H{\"o}fner}, \&
  {Aringer}}]{Nowotny2010}
{Nowotny}, W., {H{\"o}fner}, S., \& {Aringer}, B. 2010, \aap, 514, A35

\bibitem[{{Rau} {et~al.}(2017){Rau}, {Hron}, {Paladini}, {Aringer}, {Eriksson},
  {Marigo}, {Nowotny}, \& {Grellmann}}]{Rau2017}
{Rau}, G., {Hron}, J., {Paladini}, C., {et~al.} 2017, \aap, 600, A92

\bibitem[{{Rau} {et~al.}(2015){Rau}, {Paladini}, {Hron}, {Aringer},
  {Groenewegen}, \& {Nowotny}}]{Rau2015}
{Rau}, G., {Paladini}, C., {Hron}, J., {et~al.} 2015, \aap, 583, A106

\bibitem[{{Samus'} {et~al.}(2017){Samus'}, {Kazarovets}, {Durlevich},
  {Kireeva}, \& {Pastukhova}}]{Samus2017}
{Samus'}, N.~N., {Kazarovets}, E.~V., {Durlevich}, O.~V., {Kireeva}, N.~N., \&
  {Pastukhova}, E.~N. 2017, Astronomy Reports, 61, 80

\bibitem[{{Thompson} {et~al.}(2002){Thompson}, {Creech-Eakman}, \& {van
  Belle}}]{Thompson2002}
{Thompson}, R.~R., {Creech-Eakman}, M.~J., \& {van Belle}, G.~T. 2002, \apj,
  577, 447

\bibitem[{{Wachter} {et~al.}(2002){Wachter}, {Schr{\"o}der}, {Winters},
  {Arndt}, \& {Sedlmayr}}]{Wachter2002}
{Wachter}, A., {Schr{\"o}der}, K.-P., {Winters}, J.~M., {Arndt}, T.~U., \&
  {Sedlmayr}, E. 2002, \aap, 384, 452

\bibitem[{{Whitelock} {et~al.}(2008){Whitelock}, {Feast}, \& {Van
  Leeuwen}}]{Whitelock2008}
{Whitelock}, P.~A., {Feast}, M.~W., \& {Van Leeuwen}, F. 2008, \mnras, 386, 313

\bibitem[{{Wittkowski} {et~al.}(2008){Wittkowski}, {Boboltz}, {Driebe}, {Le
  Bouquin}, {Millour}, {Ohnaka}, \& {Scholz}}]{Wittkowski2008}
{Wittkowski}, M., {Boboltz}, D.~A., {Driebe}, T., {et~al.} 2008, \aap, 479, L21

\bibitem[{{Wittkowski} {et~al.}(2011){Wittkowski}, {Boboltz}, {Ireland},
  {Karovicova}, {Ohnaka}, {Scholz}, {van Wyk}, {Whitelock}, {Wood}, \&
  {Zijlstra}}]{Wittkowski2011}
{Wittkowski}, M., {Boboltz}, D.~A., {Ireland}, M., {et~al.} 2011, \aap, 532, L7

\bibitem[{{Wittkowski} {et~al.}(2016){Wittkowski}, {Chiavassa}, {Freytag},
  {Scholz}, {H{\"o}fner}, {Karovicova}, \& {Whitelock}}]{Wittkowski2016}
{Wittkowski}, M., {Chiavassa}, A., {Freytag}, B., {et~al.} 2016, \aap, 587, A12

\bibitem[{{Woodruff} {et~al.}(2009){Woodruff}, {Ireland}, {Tuthill}, {Monnier},
  {Bedding}, {Danchi}, {Scholz}, {Townes}, \& {Wood}}]{Woodruff2009}
{Woodruff}, H.~C., {Ireland}, M.~J., {Tuthill}, P.~G., {et~al.} 2009, \apj,
  691, 1328

\end{thebibliography}
\begin{acknowledgements}
Michael Scholz was a co-I of this study. He unfortunately passed away 
before seeing the final results. We are grateful for many years of his 
inspiring work, dedicated collaboration, and friendship.
GR thanks ESO for supporting her visit at ESO during her work on 
the GRAVITY data.
This research has made use of the Jean-Marie Mariotti 
Center \texttt{SearchCal} service
(\url{http://www.jmmc.fr/searchcal})
co-developed by LAGRANGE and IPAG; of CDS Astronomical Databases 
AFOEV, SIMBAD, and VIZIER
(\url{http://cdsweb.u-strasbg.fr})
and of NASA's Astrophysics Data System. We acknowledge with thanks the 
variable star observations from the AAVSO International Database 
contributed by observers worldwide and used in this
research. 
This work has made use of data from the European Space Agency (ESA) mission
{\it Gaia} (\url{https://www.cosmos.esa.int/gaia}), processed by the {\it Gaia}
Data Processing and Analysis Consortium (DPAC,
\url{https://www.cosmos.esa.int/web/gaia/dpac/consortium}). Funding for the DPAC
has been provided by national institutions, in particular the institutions
participating in the {\it Gaia} Multilateral Agreement.
\end{acknowledgements}

\clearpage
\begin{appendix}
\section{Additional material}
\begin{figure}
\resizebox{\hsize}{!}{\includegraphics{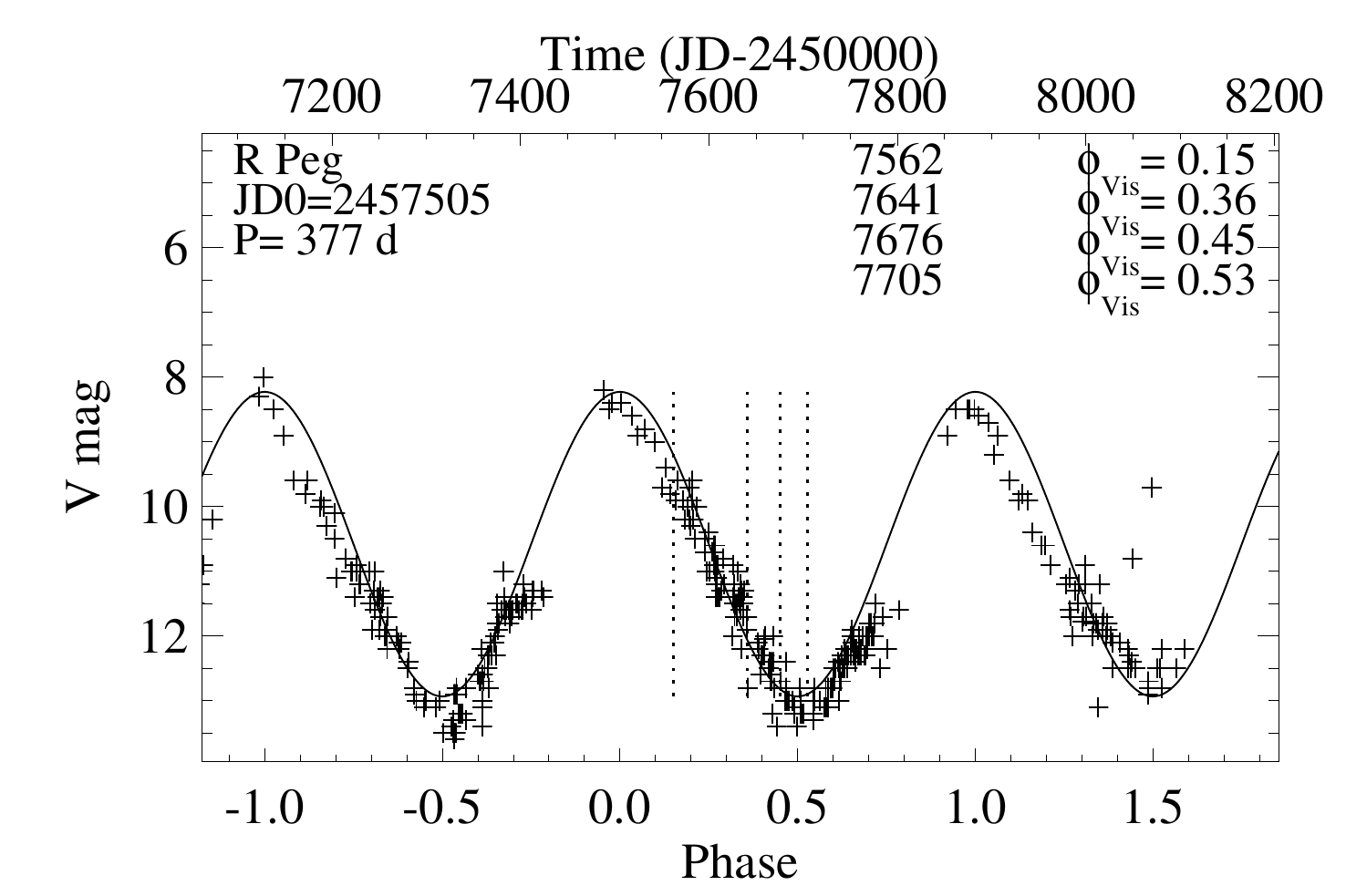}}
\caption{Visual lightcurve of R~Peg based on the AAVSO and
AFOEV databases. The solid line represents a sine fit to the
ten most recent cycles. The dashed vertical lines denote the
epochs of our observations.}
\label{fig:lightcurve}
\end{figure}

\begin{figure}
\resizebox{\hsize}{!}{\includegraphics{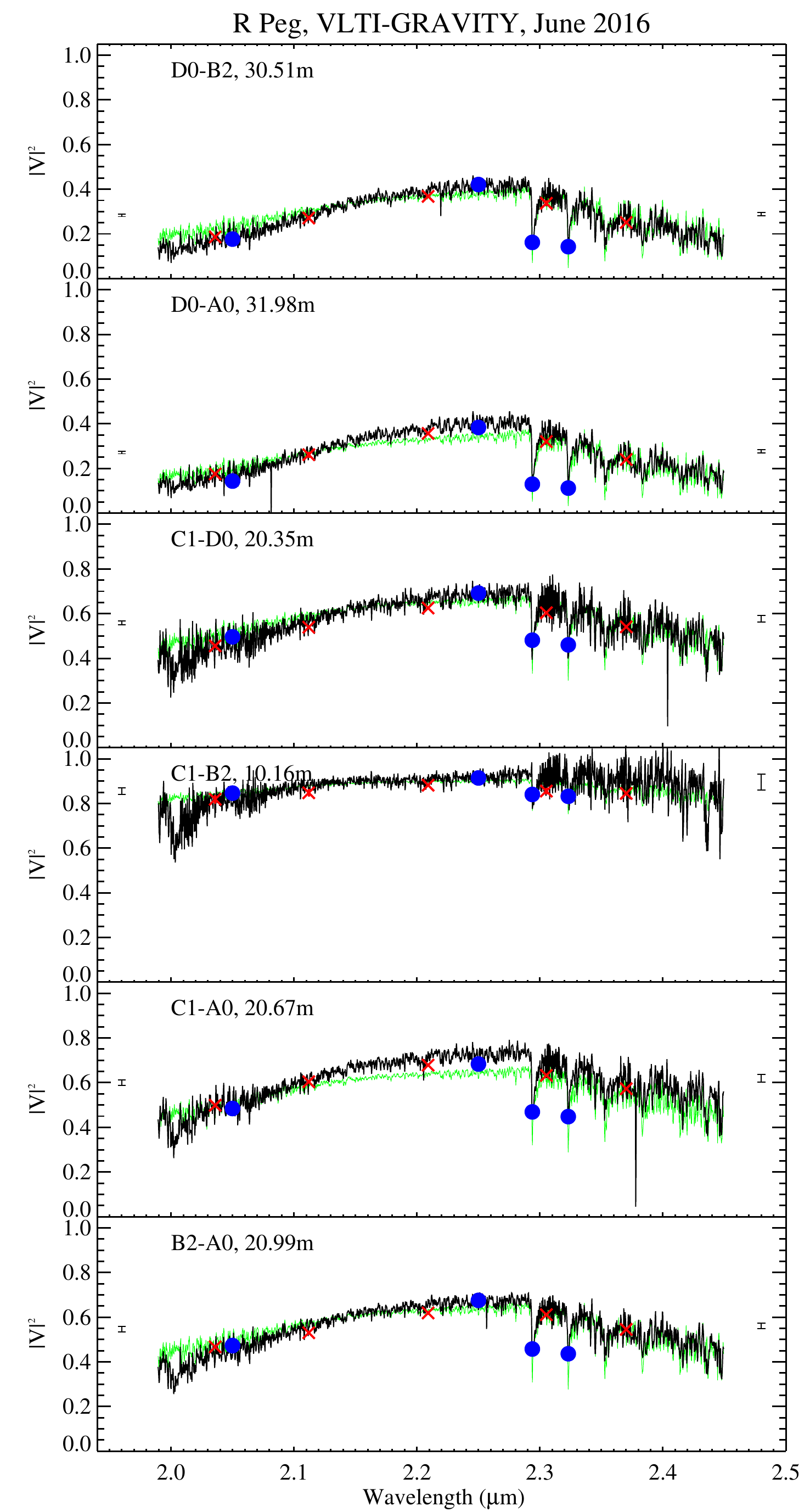}}
\caption{R~Peg visibility data obtained in June 2016. The black lines denote 
the squared visibility amplitudes measured on the science camera. The
errors are indicated by mean squared visibility amplitudes and their errors
on each side of the plot for the first and second half of the wavelength
interval. The red points with error bars denote the squared 
visibility amplitudes measured on the fringe tracker camera with 
low spectral resolution. The blue dots denote synthetic values
based on UD fits at wavelengths of 2.05\,$\mu$m (water), 
2.25\,$\mu$m (near-continuum), 2.29\,$\mu$m (CO 2-0), 
and 2.32\,$\mu$m (CO 3-1). The green lines denote a fit of a 
CODEX model atmosphere with the parameters 
listed in Table~\protect\ref{tab:fitcodex}.}
\label{fig:visjun}
\end{figure}

\begin{figure}
\resizebox{\hsize}{!}{\includegraphics{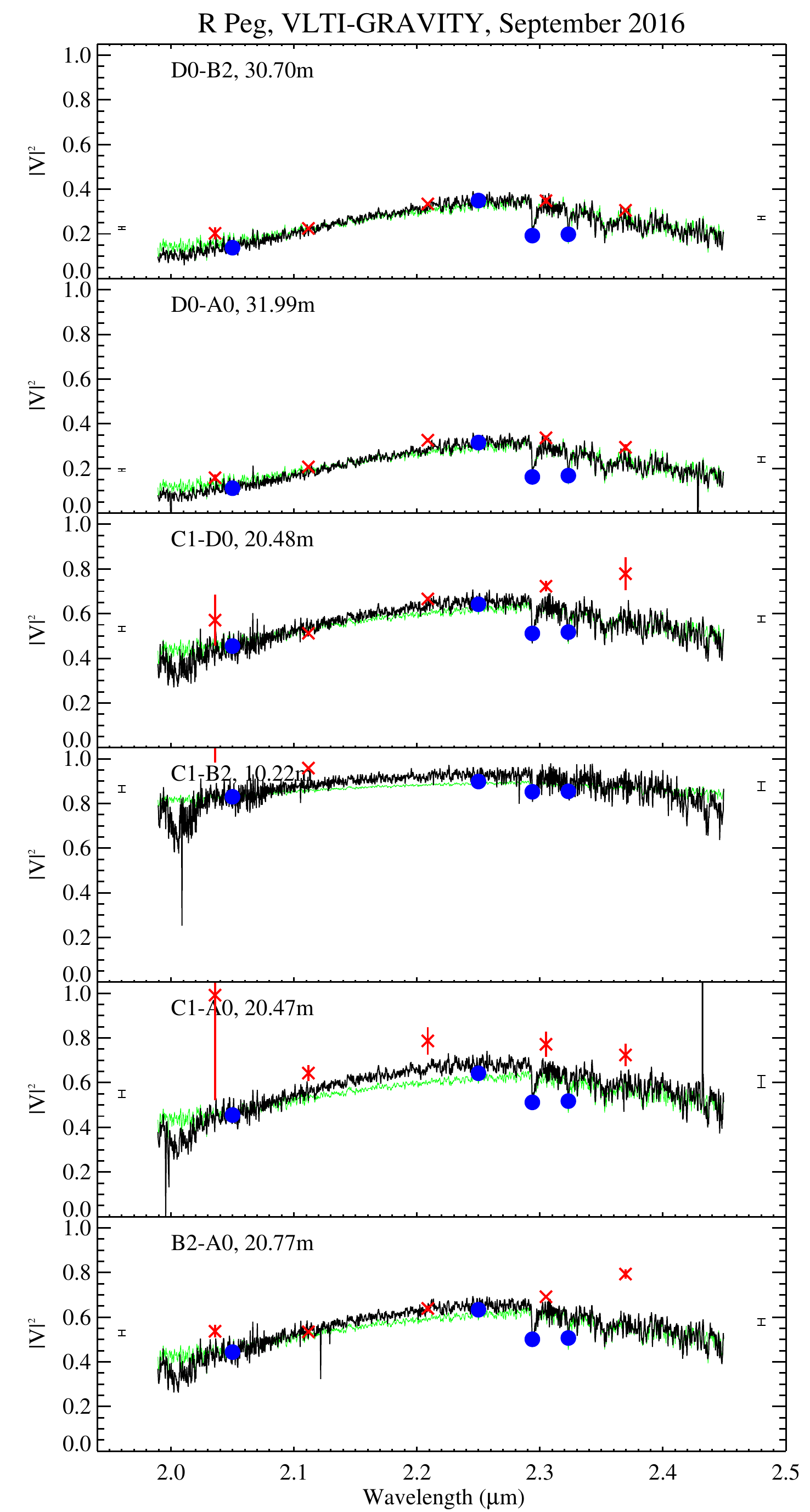}}
\caption{As in Fig.~\protect\ref{fig:visjun}, but for the data 
obtained in September.}
\label{fig:vissep}
\end{figure}

\begin{figure}
\resizebox{\hsize}{!}{\includegraphics{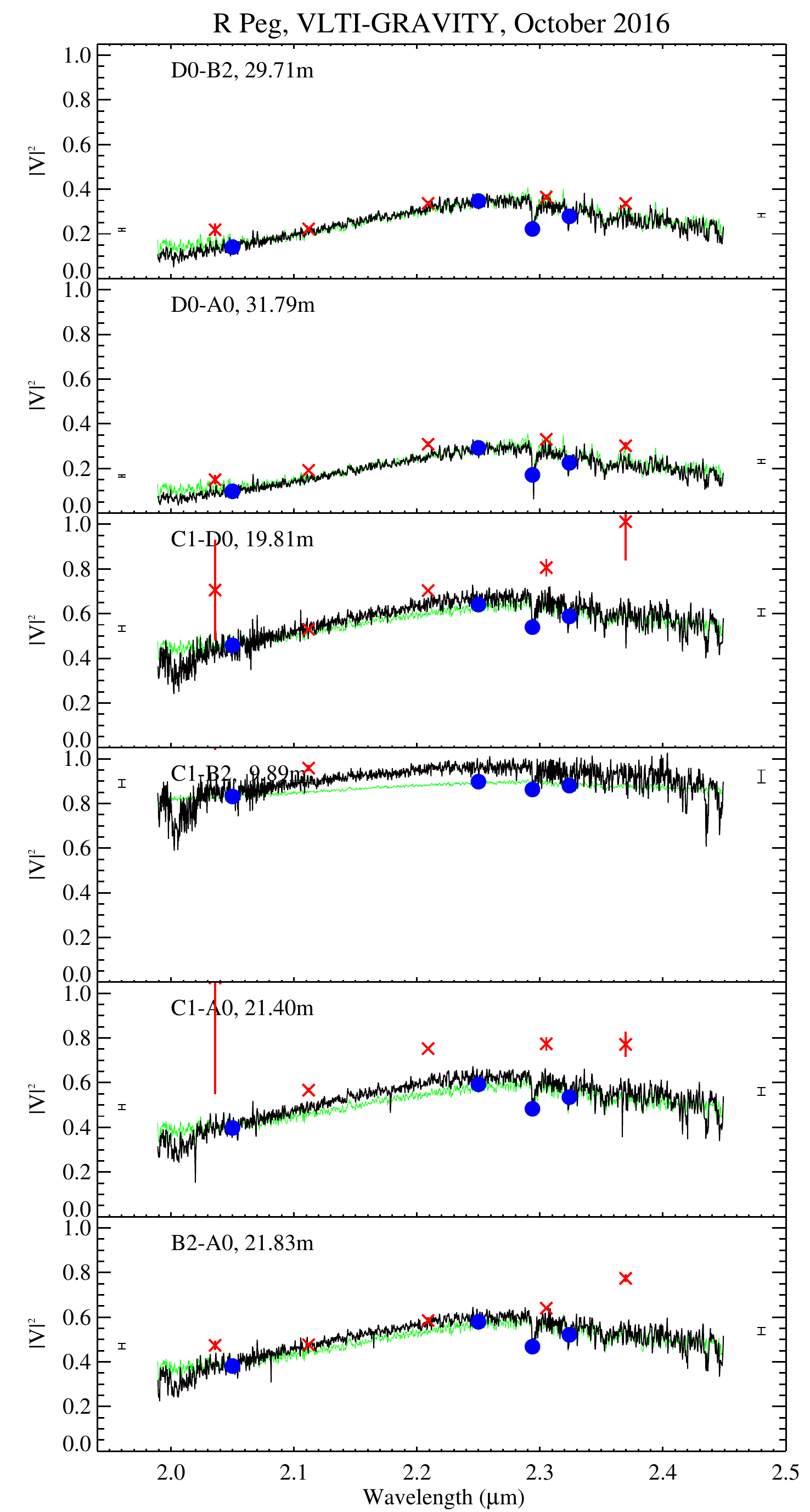}}
\caption{As in Fig.~\protect\ref{fig:visjun}, but for the data
obtained in October.}
\label{fig:visoct}
\end{figure}

\begin{figure}
\resizebox{\hsize}{!}{\includegraphics{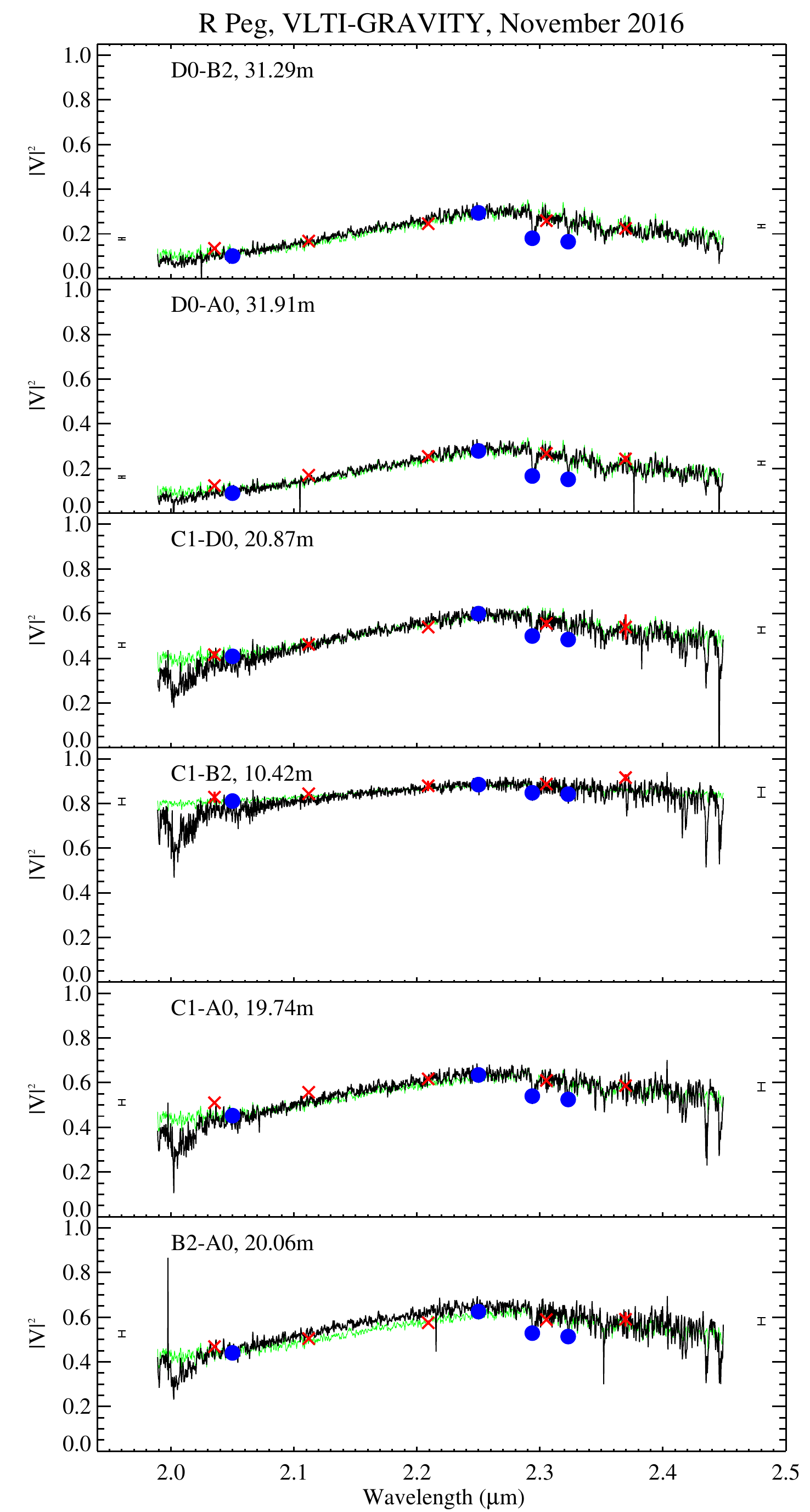}}
\caption{As in Fig.~\protect\ref{fig:visjun}, but for the data
obtained in November.}
\label{fig:visnov}
\end{figure}

\begin{figure}
\resizebox{\hsize}{!}{\includegraphics{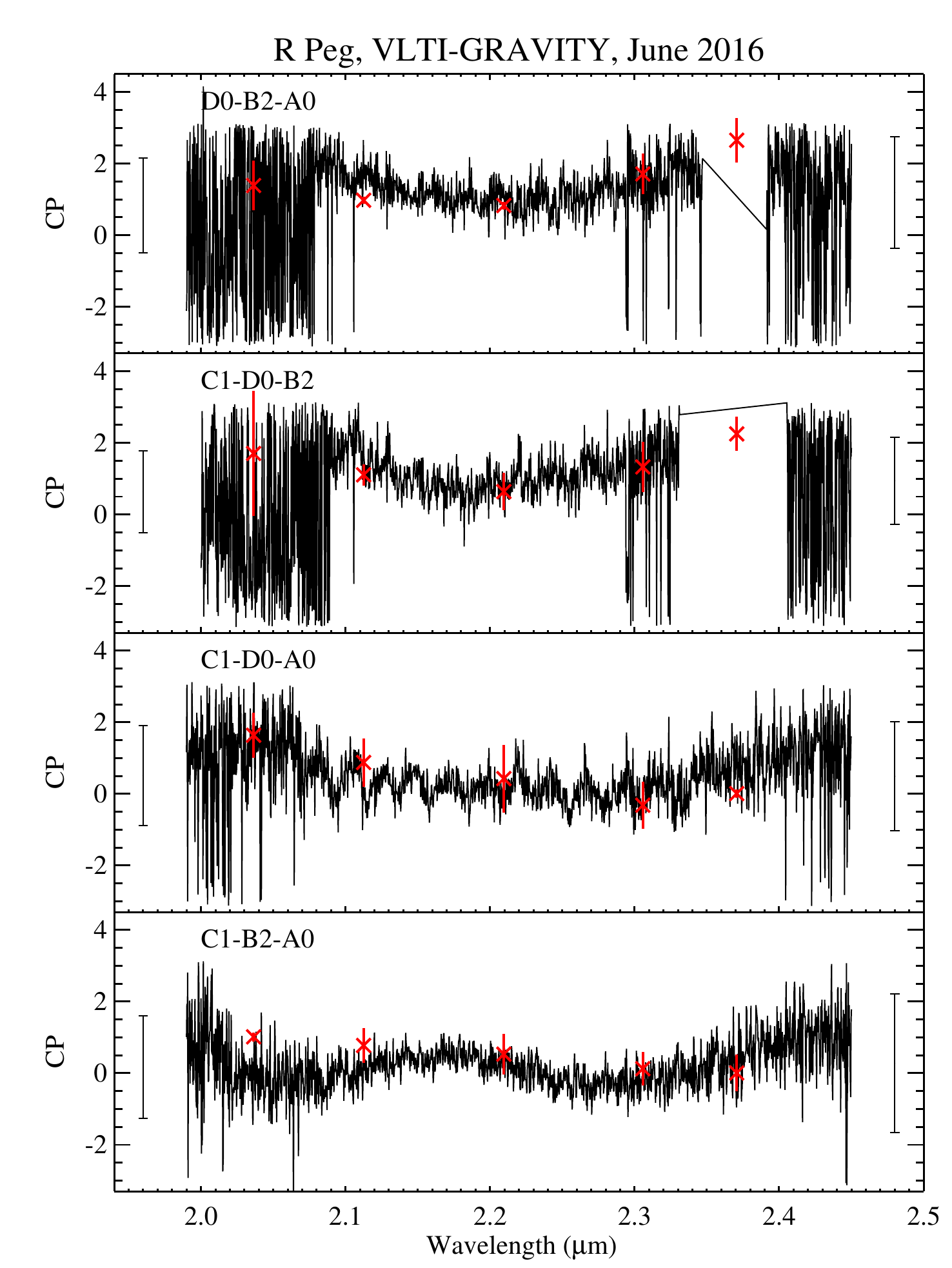}}
\caption{R~Peg closure phase obtained in June 2016. The black lines denote 
the closure phases measured on the science camera. The
errors are indicated by mean values and their errors
on each side of the plot for the first and second half of the wavelength
interval. The red points with error bars denote the closure phases
measured on the fringe tracker camera with 
low spectral resolution.}
\label{fig:cpjun}
\end{figure}

\begin{figure}
\resizebox{\hsize}{!}{\includegraphics{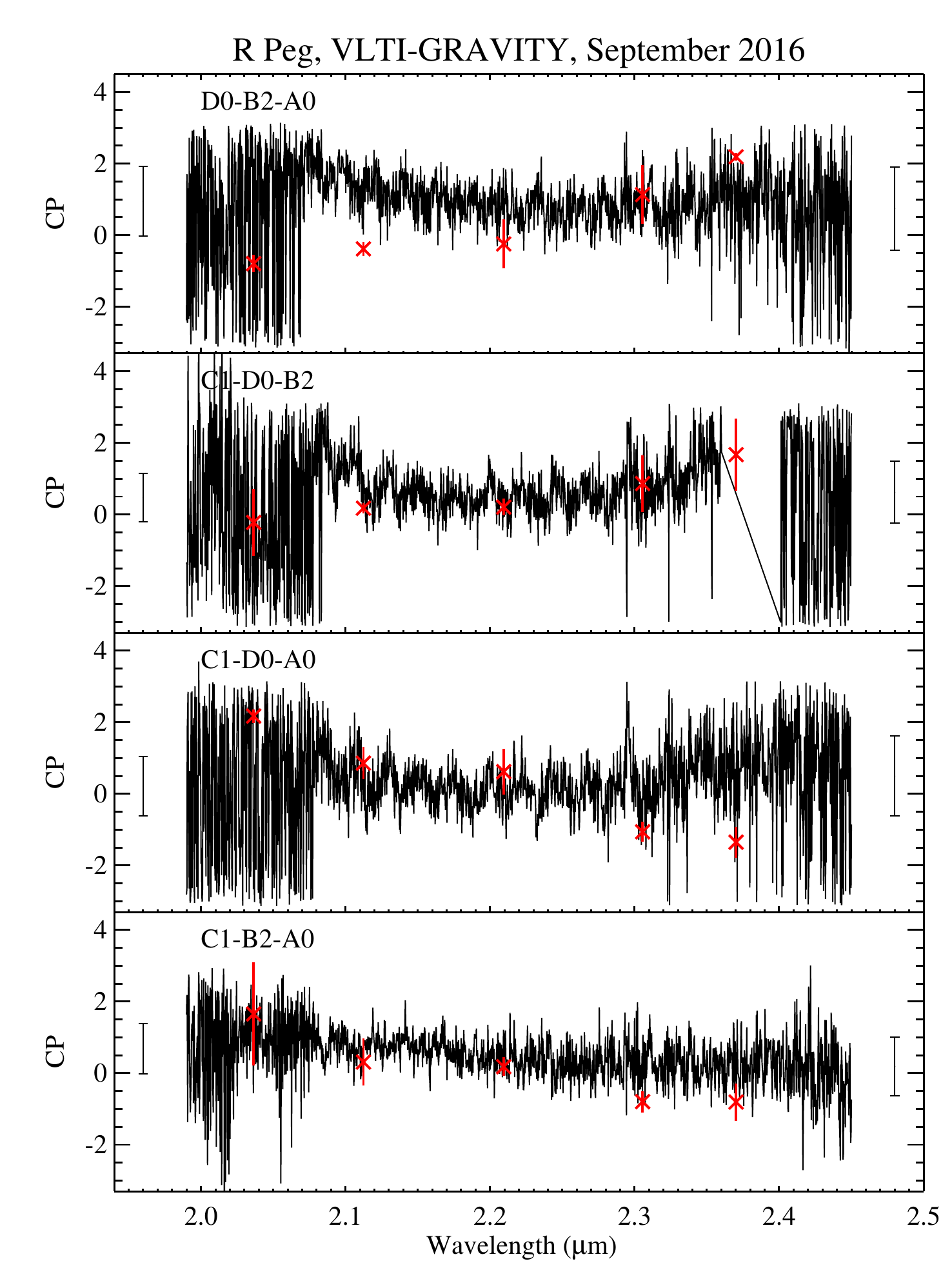}}
\caption{As in Fig.~\protect\ref{fig:cpjun}, but for the data
obtained in September.}
\label{fig:cpsep}
\end{figure}

\begin{figure}
\resizebox{\hsize}{!}{\includegraphics{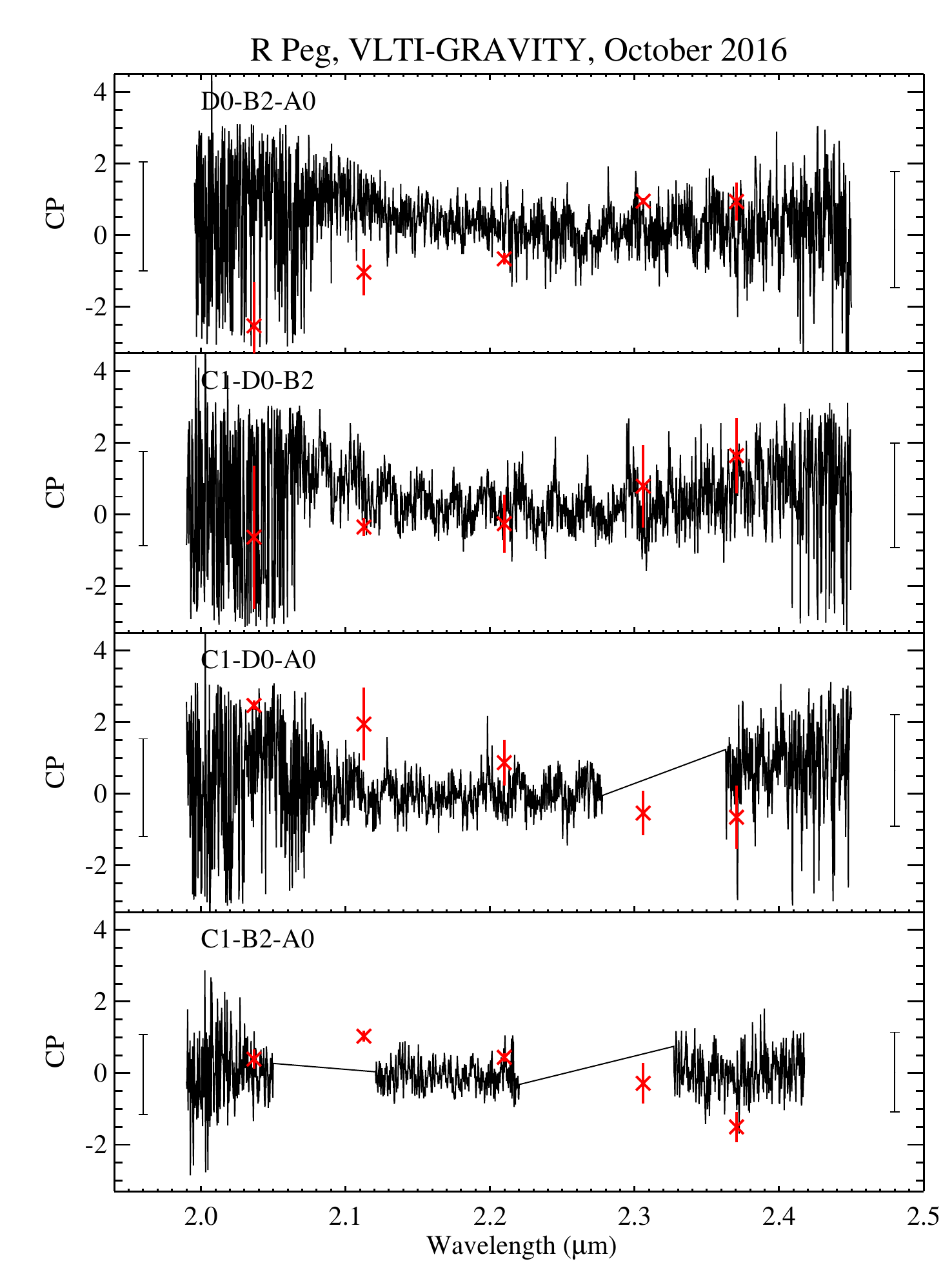}}
\caption{As in Fig.~\protect\ref{fig:cpjun}, but for the data
obtained in October.}
\label{fig:cpoct}
\end{figure}

\begin{figure}
\resizebox{\hsize}{!}{\includegraphics{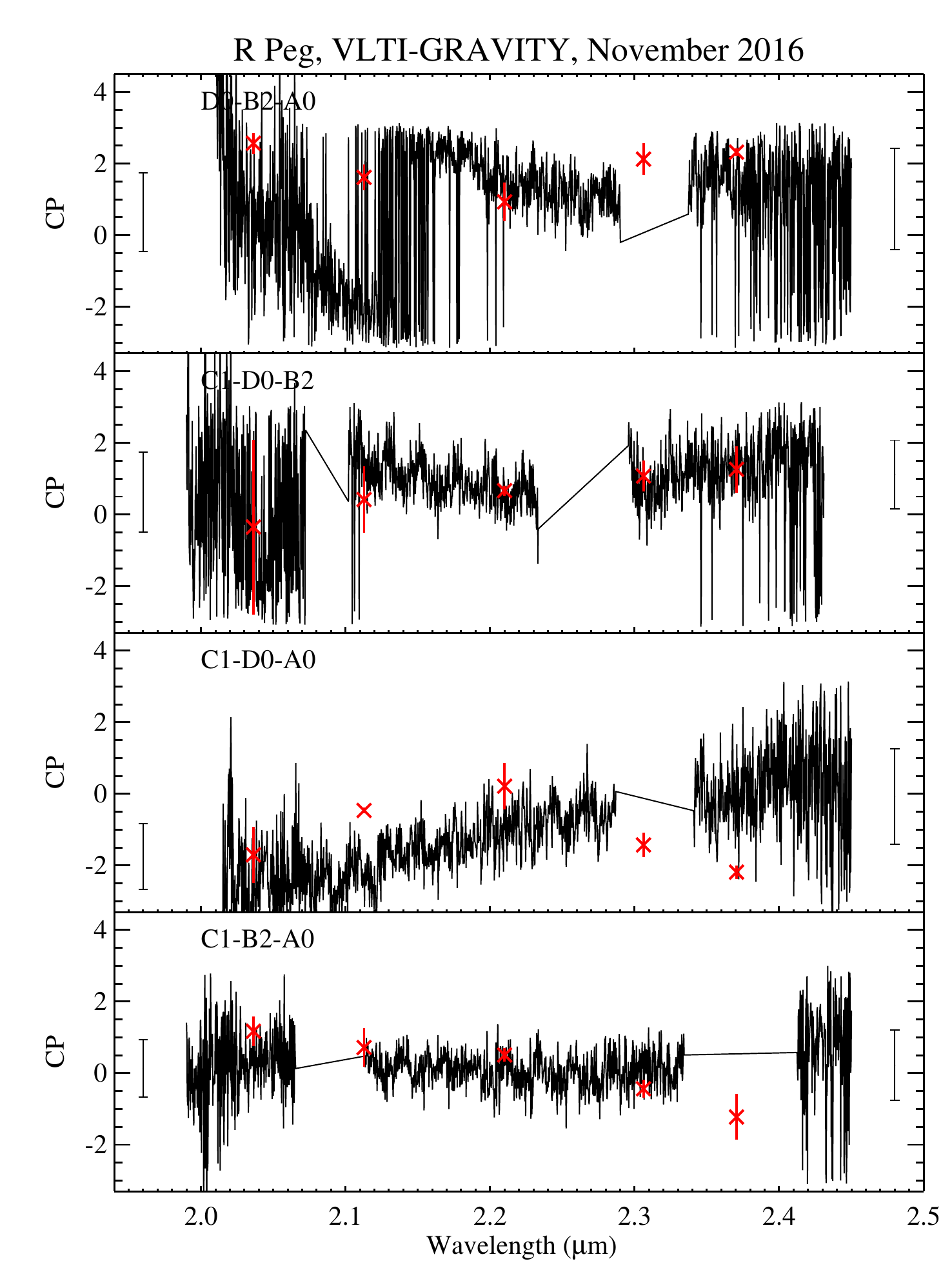}}
\caption{As Fig.~\protect\ref{fig:cpjun}, but for the data
obtained in November.}
\label{fig:cpnov}
\end{figure}

Here we provide additional material. Figure \ref{fig:lightcurve}
shows the visual lightcurve of R~Peg based on the AAVSO and
AFOEV databases for the last few cycles together with a sine fit to the
ten most recent cycles. Table~\ref{tab:obslog} details the log of our
R~Peg observations and Table~\ref{tab:calibrators} provides the
adopted properties of the calibrators. Figures \ref{fig:visjun}--\ref{fig:cpnov} show all the obtained visibility and closure phase data.
The plots of the visibility also include predictions by a series
of CODEX dynamic model atmospheres that best reproduce the observations.
\begin{table}
\caption{\label{tab:obslog} Log of our R~Peg observations.} 
\centering
\begin{tabular}{lllll}
\hline\hline
Date & JD\tablefootmark{a} & Proj. baseline  & Calibr.  &  $\phi_{V}$\\
  &   & lengths [m]\tablefootmark{b}    &            &            \\
\hline\hline
2016-06-23 & 7562.9 & 10.2/20.4/20.7/     & 36~Peg        & 0.15 \\
           &        & 21.0/30.5/32.0      &               &      \\[1ex]
2016-09-10 & 7641.7 & 10.2/20.5/20.5/     & 36~Peg,       & 0.36 \\
           &        & 20.8/30.7/32.0      & $\gamma$~Psc  &      \\[1ex]
2016-10-15 & 7676.6 &  9.9/19.8/21.4/     & HR~1452,      & 0.45 \\
           &        &  21.8/29.7/31.8     & HD~36710      &      \\[1ex]
2016-11-13 & 7705.6 & 10.4/19.7/20.1/     & HR~1452,      & 0.53 \\
           &        &  20.9/31.3/31.9     & HD~36710      &      \\
\hline
\end{tabular}\\
\tablefoot{
\tablefootmark{a}{JD-2 450 000.}
\tablefootmark{b}{All observations used the baseline configuration 
A0-B2-C1-D0 with ground baselines 11.3\,m, 22.6\,m, 22.6\,m, 25.3\,m, 
32.0\,m, 33.9\,m.}}
\end{table}
\begin{table}
\caption{\label{tab:calibrators} Properties of the interferometric 
calibrators\tablefootmark{a}.}
\centering
\begin{tabular}{lll}
\hline\hline
Target & Spectral type & Angular $K$-band \\
 &  & diameter [mas] \\
\hline
HD~1452      & F5\,V   & 0.233 $\pm$ 0.033 \\
HD~36710     & K7\,III & 1.217 $\pm$ 0.080 \\
36~Peg       & K6\,III & 2.305 $\pm$ 0.223 \\
$\gamma$~Psc & K0\,III & 2.477 $\pm$ 0.229 \\
\hline
\end{tabular}
\tablefoot{
\tablefootmark{a}{Based on \citet{Chelli2016}.}
}
\end{table}
\end{appendix}
\end{document}